\documentclass[aps,pra,twocolumn,showpacs,
amssymb,amsmath]{revtex4}
\usepackage{color}
\usepackage{graphicx}
\usepackage{dcolumn}
\usepackage{bm}

\begin{document}
\title{Characteristics of Chiral Anomaly in View of Various Applications}

\author{Kazuo Fujikawa}
\affiliation{Quantum Hadron Physics Laboratory, RIKEN Nishina Center,\\
Wako 351-0198, Japan}

\begin{abstract}
In view of the recent applications of chiral anomaly to various fields beyond particle physics, we discuss some basic aspects of chiral anomaly
which may help deepen our understanding of chiral anomaly in particle physics also. It is first shown that Berry's phase (and its generalization)  for the Weyl model $H
=v_{F} \vec{\sigma}\cdot \vec{p}(t)$ assumes a monopole form at the exact adiabatic limit but deviates from it off the adiabatic limit and vanishes in  the high frequency limit of the Fourier transform of $\vec{p}(t)$ for bounded $|\vec{p}(t)|$. An effective action, which is consistent with the non-adiabatic limit of Berry's phase,  combined with the  Bjorken-Johnson-Low prescription gives normal equal-time space-time commutators and no chiral anomaly. In contrast, an effective action with a monopole at the origin of the momentum space, which describes Berry's phase in the precise adiabatic limit but fails off the adiabatic limit, gives anomalous space-time commutators and a covariant anomaly to the gauge current. We regard this anomaly as an artifact of the postulated monopole and not a consequence of Berry's phase. As for the recent application of the chiral anomaly to the description of effective Weyl fermions in condensed matter and nuclear physics, which is closely related to the formulation of lattice chiral fermions, we point out that the chiral anomaly for each species doubler separately vanishes for a finite lattice spacing, contrary to the common assumption. Instead a general form of pair creation associated with the spectral flow for the Dirac sea with finite depth takes place. This view is supported by the Ginsparg-Wilson fermion, which defines a single Weyl fermion  without doublers on the lattice and  gives a well-defined index (anomaly) even for a finite lattice spacing.   A different use of anomaly in analogy to PCAC is also mentioned, which could lead to an effect without fermion number non-conservation.  

\end{abstract}


\maketitle
\section{Introduction}
In the treatment of topological properties in condensed matter physics, one often uses adiabatic Berry's phase~\cite{berry}  induced at the crossing point of two levels in the band structure.  See~\cite{nagaosa} for a  review of topological effects in condensed matter physics.
Chiral anomaly, which was established in particle theory~\cite{anomaly1,anomaly2,anomaly3,bertlmann}, has also been used to elucidate the properties of ``Weyl fermions'' in condensed matter and nuclear physics.  See, for example, \cite{nielsen2, fukushima, zyuzin, hosur} and references therein. 

It is clear from its original derivation~\cite{berry}  that the monopole-type behavior of Berry's phase is valid only in the precise adiabatic limit~\cite{simon}. On the other hand, chiral anomaly is believed to be short distance effects~\cite{wilson} and in fact  only the high frequency components of  fermion variables are essential in the (Euclidean) evaluation of chiral anomaly~\cite{fujikawa-suzuki}, and thus the relation~\cite{anomaly1,anomaly2}
\begin{eqnarray}\label{identity1}
&&\partial_{\mu}\left(\bar{\psi}\gamma^{\mu}\gamma_{5}\psi\right)=
2im\bar{\psi}\gamma_{5}\psi+\frac{e^{2}}{2\pi^{2}}\vec{E}\cdot\vec{B}
\end{eqnarray}
holds for the fundamental electron in condensed matter in an arbitrary small domain of space-time (such as the tangent space of a curved space) independently of frequencies carried by the gauge field, which may include the Coulomb potential provided by surrounding charged particles in addition to external field $A_{\mu}$. 

It has been recently argued, relying on some preceding analyses~\cite{niu, duval}, that a ``kinematic'' derivation of  chiral anomaly from Berry's phase is possible in an effective theoretical model~\cite{yamamoto}. If this derivation, which is based on anomalous commutation relations of space-time variables induced by Berry's phase, is valid it would open up completely new perspective for the subject of quantum anomalies. The purpose of the present paper is firstly to examine the foundation of this kinematic derivation and then secondly to discuss some basic issues related to the applications of the chiral anomaly in condensed matter and nuclear physics.

We first show that Berry's original model is exactly solved if the time dependence of parameters are suitably chosen. This solution  enables us to study Berry's phase in the non-adiabatic and non-topological domain and shows that,  in particular, Berry's phase (its non-adiabatic generalization~\cite{aharonov}) vanishes in the non-adiabatic limit when measured in a natural manner.  Using this knowledge of the non-adiabatic behavior, it is shown that the mere presence of a Weyl fermion does not induce anomalous equal-time commutation relations of space-time variables  in the Bjorken-Johnson-Low prescription.
In contrast, an effective action with a postulated point-like monopole at the origin of the momentum space, which describes Berry's phase in the precise adiabatic limit but fails off the adiabatic limit, gives rise to anomalous space-time commutators and  a covariant form of anomaly to the gauge current~\cite{yamamoto}. It is well-known that the magnetic field modifies commutators such as in the Landau level, and thus the appearance of anomalous space-time commutators for an assumed monopole in the momentum space is not surprising. We regard the anomaly thus obtained as an artifact of the postulated monopole and not a consequence of Berry's phase.  

In the application of the chiral anomaly in condensed matter physics and related fields, the species doublers are treated as physical objects unlike in conventional lattice gauge theory where they are unphysical nuisances. We point out that the chiral anomaly generated by $\gamma_{5}$ for species doublers vanishes for each doubler separately in the chiral symmetric lattice gauge theory with a finite lattice spacing, contrary to the common assumption. Instead the general form of the pair creation associated with the spectral flow appears; it is emphasized that a picture of spectral flow is drastically changed if the Dirac sea has a finite depth. This view is supported by the Ginsparg-Wilson fermion that describes a Weyl fermion on the lattice without species doublers and gives a well-defined index related to the chiral anomaly even for a finite lattice spacing. 

The present study is motivated by the recent excitement in the subject of effective ``Weyl fermions'' in condensed matter and nuclear physics~\cite{nagaosa, nielsen2, fukushima, zyuzin, hosur, niu, duval, yamamoto}, but we believe that it contains physical and technical aspects which will interest the wider audience. 

\section{Berry's phase} 
 
We here discuss some general properties of Berry's phase, which are relevant to the analyses in the present paper,  using a simple soluble model. 
 The precise adiabatic Berry's phase is defined only in an ideal mathematical limit.  
 We thus adopt the exact non-adiabatic phase (holonomy) in the manner of Aharonov and Anandan for the general Schroedinger problem~\cite{aharonov}, which contains enough freedom to describe a possible genuine monopole,  as our definition of Berry's phase.  The adiabatic Berry's phase is naturally defined from the non-adiabatic phase in the adiabatic limit. The generic term ``geometric phase'' is thus more appropriate, but we follow the common practice and use the term   
``Berry's phase'' except for the case where it is appropriate to make a distinction. 

To analyze the behavior of Berry's phase away from the precise adiabatic limit quantitatively, we discuss an exactly solvable model which is defined by 
\begin{eqnarray}
H=-\hbar\mu\vec{B}(t)\vec{\sigma}
\end{eqnarray}
with 
\begin{eqnarray}\label{magnetic-field}
\vec{B}(t)=B(\sin\theta\cos\varphi(t), 
\sin\theta\sin\varphi(t),\cos\theta )
\end{eqnarray}
where $\varphi(t)=\omega t$ with constant $\omega$, $B$ and 
$\theta$, and $\vec{\sigma}$ stand for Pauli matrices.  This model is identical to the original model analyzed by Berry~\cite{berry}, except for the choice of specific time dependence (or independence) of parameters so that we can solve the model exactly. 
The  exact solution of the Schr\"{o}dinger equation,
$i\hbar\partial_{t}\psi(t)=H\psi(t)$,
 is given by
\begin{eqnarray}\label{exact}
\psi_{\pm}(t)
&=&w_{\pm}(t)\exp\left[-\frac{i}{\hbar}\int_{0}^{t}dt^{\prime}
w_{\pm}^{\dagger}(t^{\prime})\big(H
-i\hbar\partial_{t^{\prime}}\big)w_{\pm}(t^{\prime})\right]
\end{eqnarray}
where
\begin{eqnarray}\label{eigen-1}
w_{+}(t)&=&\left(\begin{array}{c}
            \cos\frac{1}{2}(\theta-\alpha) e^{-i\varphi(t)}\\
            \sin\frac{1}{2}(\theta-\alpha)
            \end{array}\right), \nonumber\\ 
w_{-}(t)&=&\left(\begin{array}{c}
            \sin\frac{1}{2}(\theta-\alpha) e^{-i\varphi(t)}\\
            -\cos\frac{1}{2}(\theta-\alpha)
            \end{array}\right).
\end{eqnarray}
This solution is confirmed by evaluating
\begin{eqnarray}
&&i\hbar\partial_{t}\psi(t)\nonumber\\
&&=\{ i\hbar\partial_{t}w_{\pm}(t)+w_{\pm}(t)[w_{\pm}^{\dagger}(t)\big(H
-i\hbar\partial_{t}\big)w_{\pm}(t)]\}\nonumber\\
&&\times\exp\left[-\frac{i}{\hbar}\int_{0}^{t}dt^{\prime}
w_{\pm}^{\dagger}(t^{\prime})\big(H
-i\hbar\partial_{t^{\prime}}\big)w_{\pm}(t^{\prime})\right]\nonumber\\
&&=\{ i\hbar\partial_{t}w_{\pm}(t)+w_{\pm}(t)[w_{\pm}^{\dagger}(t)\big(H
-i\hbar\partial_{t}\big)w_{\pm}(t)]\nonumber\\
&&+w_{\mp}(t)[w_{\mp}^{\dagger}(t)\big(H
-i\hbar\partial_{t}\big)w_{\pm}(t)]\}\nonumber\\
&&\times\exp\left[-\frac{i}{\hbar}\int_{0}^{t}dt^{\prime}
w_{\pm}^{\dagger}(t^{\prime})\big(H
-i\hbar\partial_{t^{\prime}}\big)w_{\pm}(t^{\prime})\right]\nonumber\\
&&=H\psi(t)
\end{eqnarray}  
where we used   
$w_{\mp}^{\dagger}[H-i\hbar\partial_{t}]w_{\pm}=0$
 and the completeness relation $w_{+}w_{+}^{\dagger}+w_{-}w_{-}^{\dagger}=1$. 

The parameter $\alpha$ is given by
\begin{eqnarray}\label{parameter-alpha1}
\eta \sin\alpha=\sin(\theta-\alpha),
\end{eqnarray}
or equivalently, $\tan\alpha = \sin\theta/(\eta+
\cos\theta)$,
with the parameter $\eta$ defined by 
\begin{eqnarray}\label{eta}
\eta \equiv 2\hbar\mu B/\hbar\omega = \mu BT/\pi
\end{eqnarray}
and the period $T=2\pi/\omega$.
The parameter $\eta\rightarrow \infty$ implies the adiabatic limit and $\eta\rightarrow 0$ implies the non-adiabatic limit.  

The {\em exact} extra phase factor in \eqref{exact} 
for one period of motion 
\begin{eqnarray}\label{exact-formula}
&&\exp\left[-\frac{i}{\hbar}\int_{0}^{T}dt
w_{\pm}^{\dagger}(t)\big(-i\hbar\partial_{t}\big)w_{\pm}(t)\right]\nonumber\\
&=&\exp\left[-i\pi(1\mp\cos(\theta-\alpha))\right]
\end{eqnarray}
defines Berry's phase 
\begin{eqnarray}\label{phaseformula}
\Omega_{\pm}&\equiv& 2\oint
\vec{{\cal A}}_{\pm}(\vec{B})\frac{d\vec{B}}{d t}dt\nonumber\\
&=&2\pi\left(1\mp\cos(\theta-\alpha)\right)
\end{eqnarray}
where Berry's connection is defined by
\begin{eqnarray}\label{connection}
\vec{{\cal A}}_{\pm}(\vec{B})=w_{\pm}^{\dagger}(t)\big(-i\partial_{\vec{B}}\big)w_{\pm}(t).
\end{eqnarray}
 We tentatively normalize $\Omega$ in this section to agree with the solid angle subtended by the moving spin. 
It is known~\cite{non-adiabatic} that our phase agrees with the non-adiabatic phase in the manner of Aharonov and Anandan. The exact expression of $\Omega_{\pm}$ in \eqref{phaseformula}
as it stands is {\em not topological} because of the presence of $\alpha$,
which modifies the ``magnetic flux'' $\Omega_{+}$, for example, from the monopole value
\begin{eqnarray}
\Omega_{{\rm mono}}=2\pi(1-\cos\theta).
\end{eqnarray}
The topology of Berry's phase is thus only approximate for the generic situation $\eta<\infty$ for which $\alpha\neq 0$.
For the precise adiabatic limit $T \rightarrow \infty$~\cite{simon} (and thus $\eta \rightarrow \infty$) for which
$\alpha \rightarrow 0$ in \eqref{parameter-alpha1}, we have 
$\Omega_{\pm} \rightarrow 2\pi(1\mp \cos\theta)$,
namely, approaches the phase generated by a spurious monopole located at the origin of the parameter space $\vec{B}=0$. The limit $T\rightarrow \infty$ implies that one cannot define the winding number since it takes an infinite amount of time to make one turn; this trivial topology is also seen by the vanishing of Berry's phase by deforming it smoothly to the non-adiabatic limit using the exact solution, as shown below.

In the  non-adiabatic limit $\eta=\mu BT/\pi \rightarrow 0$, which includes the cases  $B\rightarrow 0$ with fixed $T$ or $T\rightarrow 0$ with fixed $B$, we have $\alpha\rightarrow \theta$ in \eqref{parameter-alpha1} and thus Berry's phase $\Omega_{\pm}$ become  trivial constants, 
$\Omega_{\pm} \rightarrow 0\  {\rm or}  \ 4\pi$.
Namely, the ``magnetic monopole'' disappears for $\eta\rightarrow 0$ which includes the approach to the spurious monopole position $B\rightarrow 0$ with fixed $T$ or $T\rightarrow 0$ with fixed $B$ in the exact solution; physically, this means that the non-adiabatic phase cannot describe a genuine monopole in the non-adiabatic domain such as   $B\rightarrow 0$ with fixed $T$.  
It is emphasized that we do not assign physical reality to our ``monopole''; we define the monopole operationally, and we say that there is a monopole if one finds a well-defined phase corresponding to a monopole.

 To analyze the vanishing of the monopole more quantitatively we define the ratio $\Omega_{+}/\Omega_{{\rm mono}} =(1-\cos(\theta-\alpha))/(1-\cos\theta)$ which approaches 
$\Omega_{+}/\Omega_{{\rm mono}} =\eta^{2}\cos^{2}\frac{1}{2}\theta$
for $\eta\rightarrow 0$ since we have $\theta-\alpha\simeq \eta\sin\theta$ from \eqref{parameter-alpha1} in this limit. We thus have for the upper half sphere with $\theta=\pi/2$,
\begin{eqnarray}\label{non-adiabatic}
\Omega_{+}/\Omega_{{\rm mono}}=(1/2)\eta^{2}
\end{eqnarray}
which is an indicator how the ``monopole'' flux vanishes in the non-adiabatic limit $\eta=2\mu B/\omega \rightarrow 0$. This in particular shows that
\begin{eqnarray}\label{non-adiabatic2}
\Omega_{+}/\Omega_{{\rm mono}}\sim 1/\omega^{2}
\end{eqnarray}
for $\omega \rightarrow \infty$. (If one starts with the south pole, one may use $\Omega^{\prime}_{{\rm mono}}=2\pi(1+\cos\theta)$ and $\Omega_{-}$ and the fact that $\Omega$ is defined mod $4\pi$, then one obtains the same ratio.)    Physically, this shows that the movement of the spin does not follow the rapid movement of $\vec{B}$ or very weak $\vec{B}$; it is well known that Berry's phase is understood as the solid angle subtended by the moving spin (see, for example, \cite{non-adiabatic})
\begin{eqnarray}\label{solid-angle}
&&\psi_{\pm}^{\dagger}(t)\vec{\sigma}\psi_{\pm}(t)\nonumber\\ &=& w_{\pm}^{\dagger}(t)\vec{\sigma}w_{\pm}(t) \\
&=& \pm(\sin(\theta-\alpha)\cos\varphi,\sin(\theta-\alpha)
\sin\varphi,\cos(\theta-\alpha)).\nonumber
\end{eqnarray}

In passing we mention that, 
using the hidden-local gauge symmetry inherent in the second quantized formulation,
 one can parameterize any  solution of the Schroedinger equation
$i\hbar\partial_{t} \psi(t,\vec{x})=\hat{H}(t) \psi(t,\vec{x})$
for a general Hamiltonian $\hat{H}(t)$,   which is cyclic (i.e., periodic up to a phase factor),  in the form
\begin{eqnarray}
\psi(t,\vec{x})
&=&w(t,\vec{x})\exp\{-\frac{i}{\hbar}[\int_{0}^{t}dt \int d^{3}x  w^{\dagger}(t,\vec{x})\hat{H}(t) w(t,\vec{x})\nonumber\\
&&-\int_{0}^{t}dt \int d^{3}x w^{\dagger}(t,\vec{x}) i\hbar\partial_{t} w(t,\vec{x})]\},
\end{eqnarray}
with a suitable function $w(t,\vec{x})$ with $w(0,\vec{x})=w(T,\vec{x})$~\cite{non-adiabatic}. Our  exact solution in \eqref{exact} has this general structure.
The non-adiabatic phase (holonomy) in the manner of Aharonov and Anandan~\cite{aharonov} is then written as
\begin{eqnarray}\label{aharonov-anandan}
&&arg\{\psi(0,\vec{y})^{\dagger}\exp[\frac{i}{\hbar}\int_{0}^{T}dt \int d^{3}x \psi^{\dagger}(t,\vec{x}) i\hbar\partial_{t} \psi(t,\vec{x})]\psi(T,\vec{y})\}\nonumber\\
&& =-\frac{1}{\hbar}\int_{0}^{T}dt \int d^{3}x w^{\dagger}(t,\vec{x})(-i\hbar\partial_{t}) w(t,\vec{x}).
\end{eqnarray}
This shows that Berry's phase used in the present paper  in \eqref{connection} for the spatially independent problem with $w(t)$ is  the nonadiabatic phase~\cite{non-adiabatic}.  It is known that the non-adiabatic phase for the spin system agrees with the  solid angle subtended by the moving spin as in \eqref{solid-angle} (see, for example,~\cite{non-adiabatic}).
It is also possible to consider that our formula \eqref{exact-formula} is measuring the generally defined  geometric phase \eqref{aharonov-anandan} using the specific magnetic field \eqref{magnetic-field}.    
See also an old paper of E. Majorana~\cite{majorana}.

In application of Berry's phase to condensed matter 
physics~\cite{nagaosa, zyuzin, hosur}, one analyzes a $2\times 2$ energy matrix $h(\vec{p}(t))$ which is a truncation of band structure to the two levels crossing at $\vec{p}(t)=0$. The variable $\vec{p}(t)$ specifies the movement of the electron along the levels in the band structure; in a semi-classical picture, 
$\frac{d}{dt}\vec{p}=e[\vec{E}+\vec{v}_{F}\times \vec{B}]$.  This two-level truncation of the multi-band structure is expected to be valid at sufficiently close to the level crossing point.  It is generally expanded as  
\begin{eqnarray}\label{pseudospin}
h(\vec{p}(t))
&=&\left(\begin{array}{cc}
            y_{0}(t)&0\\
            0&y_{0}(t)
            \end{array}\right)
        + \vec{\sigma}\cdot \vec{y}(t)
\end{eqnarray}
where $\vec{\sigma}$ stand for Pauli matrices; $(y_{0}, \vec{y})$ are functions of $\vec{p}(t)$.  Ignoring the common term $y_{0}(t)$ one may consider a special (Weyl) case $\vec{y}=v_{F} \vec{p}$ with $v_{F}$ a suitable constant, and one obtains  the Hamiltonian with {\em pseudo-spin},
\begin{eqnarray}\label{hamiltonian}
H=v_{F} \vec{\sigma}\cdot \vec{p}(t)
\end{eqnarray}
 which is the case we discuss below and  
mathematically identical to the above motion of spin in a rotating magnetic field.  A more general case is analyzed in \cite{zyuzin}. The property of Berry's phase explained above in \eqref{non-adiabatic2} shows that Berry's phase vanishes in the non-adiabatic limit $\omega=2\pi/T\rightarrow  \infty$ with fixed $B$. This property is crucial in the analysis of possible anomalous commutation relations induced by Berry's phase since the limit $\omega\rightarrow  \infty$ determines the {\em equal-time} commutation relations, which imply high frequencies by the uncertainty principle,  of the variable involved in Berry's phase such as $\vec{B}(t)$ or $\vec{p}(t)$ by the  Bjorken-Johnson-Low (BJL) prescription~\cite{BJL}.

Before proceeding further, we emphasize that the notion of topology of an adiabatic phase is ``self-contradictory'' in the sense that topology implies 
its robustness against the general variation of dynamical variables and parameters, while the 
adiabatic phase is well defined only in the precise adiabatic limit. Mathematically, one may first take the precise adiabatic limit $|B| \rightarrow \infty$ with fixed $T=2\pi/\omega$ or $T \rightarrow \infty$ with fixed $|B|$, as was done by B. Simon~\cite{simon}, and examine the remaining  topological structure, namely, a monopole with
$\Omega_{{\rm mono}}=2\pi(1-\cos\theta)$ in our model; to be precise, this monopole is topological (invariant) with respect to $T$ for each $\theta$  when $|B| \rightarrow \infty$ is first taken, and topological with respect to $|B|$ for each $\theta$  when $T \rightarrow \infty$ is first taken. In comparison, a genuine monopole, if it should exist,  is topological with respect to both $(T, |B|)$ for each $\theta$.  This fact shows that the derivation of a genuine monopole from the adiabatic Berry's phase is  a very crude approximation.
\\
\ \ The parameter domain is too strongly constrained in the above mathematical  adiabatic limit~\cite{simon} to analyze equal-time commutators in our application.  We thus work with the exact nonadiabatic phase (holonomy)~\cite{aharonov} that is {\em geometric} giving a solid angle drawn by the moving spin as in \eqref{solid-angle}.  This geometric phase has enough parameters to describe a possible genuine monopole if it should be contained in the geometric phase as a subclass. {\em It should be emphasized that we do not assume the physical reality of our monopole.}
We measure the geometric phase (flux from a ``monopole'') using  Berry's connection \eqref{connection} operationally.  When the geometric phase agrees with the solid angle given by a monopole, we say that there is a monopole. If the geometric phase vanishes, we say that the monopole disappeared.  The same is true in the next section, where we measure the geometric phase using Berry's connection defined in terms of  momentum. 
\\
\ \ We have shown that the geometric phase thus defined  approaches the monopole value at the adiabatic limit, for example, $\omega\rightarrow 0$ with fixed non-vanishing $|B|$, but as soon as one is away from the exact adiabatic limit, the geometric phase deviates from the monopole value.  The solid angle which detemines the geometric phase shrinks to 0 in the non-adiabatic limit $\eta\rightarrow 0$, for example, for $|B| \rightarrow 0$ with fixed finite $\omega$ or  $\omega \rightarrow \infty$ with $|B|\leq B_{0}$ with a constant $B_{0}$ as  \eqref{solid-angle} shows. The  geometric phase  vanishes in the non-adiabatic limit, while the flux from a genuine monopole, if it should exist, does not vanish.  This limiting behavior is crucial for our analyses in the next section.

\section{Commutators in BJL  prescription}

To analyze the possible anomalous commutation relations induced by anomalies which are not recognized by naive canonical manipulations~\cite{anomaly1,anomaly2,anomaly3,bertlmann}, one needs to use a machinery which does not rely on the canonical argument. The BJL prescription  provides such a scheme and has been used extensively to analyze commutators related to anomalies~\cite{jackiw, adler, faddeev, jo}.
We emphasize that the BJL prescription has worked for all the known cases of anomalous commutators associated with anomalies. 
 
 To be explicit, we start with the time ordered correlation function of dynamical variables such as
 \begin{eqnarray} 
\int_{-\infty}^{\infty} dt e^{i\omega t}\langle T^{\star} x_{k}(t)x_{l}(0)\rangle
\end{eqnarray}
where $T^{\star}$ stands for the so-called covariant T-product,
which does not specify the precise equal-time limit $t = 0$  of the correlation function. We assume that an explicit form of the correlation function is known by the path integral evaluation, for example. The basic observation of BJL is  that, if the above correlation vanishes for $\omega\rightarrow \infty$, one can replace $T^{\star}$ by the conventional $T$ product which is assumed to specify the equal-time limit precisely.  This criterion is regarded as an analogue~\cite{fujikawa-suzuki} of Riemann-Lebesgue lemma in Fourier transform; if function $f(t)$ is smooth and well-defined around $t=0$, the large frequency limit of $\int_{-\infty}^{\infty} dt e^{i\omega t}f(t)$ vanishes. We can thus identify the canonical $T$  product from the quantity defined by $T^{\star}$ product.  

We now examine an explicit effective action for the electron 
\begin{eqnarray}\label{action}
S=\int dt[p_{k}\dot{x}_{k} +A_{k}(\vec{x},t)\dot{x}_{k} - {\cal A}_{k}(\vec{p})\dot{p}_{k} -\frac{(\vec{p}-\vec{p}_{F})^{2}}{2m}]
\end{eqnarray}
which played a central role in the analyses of anomalous commutation relations induced by ``Berry's phase'' in references~\cite{yamamoto, niu, duval}. This action is constructed to reproduce the equations of motion in the precise adiabatic limit.  $A_{k}(\vec{x},t)$ is the electromagnetic potential whose explicit time dependence is assumed to be very slow in our analysis. The Berry connection  ${\cal A}_{k}(\vec{p})$ stands for an analogue of \eqref{connection} defined by the Hamiltonian \eqref{hamiltonian}, of which expression has a well-defined monopole form only in the precise adiabatic limit $\eta= 2v_{F}|\vec{p}|/\hbar\omega\rightarrow\infty$, namely, sufficiently slow time dependence of $\vec{p}(t)$ and sizable $|\vec{p}(t)|$.  Customarily,  an {\em exact} monopole form is assumed for the general movement of dynamical variables on the premise that one applies the action only to the approximately adiabatic movement of variables ~\cite{yamamoto, niu, duval}.  We follow this procedure for a moment, but come back to a more precise re-examination of \eqref{action} later.   
We choose the form of the free Hamiltonian $(\vec{p}-\vec{p}_{F})^{2}/(2m)$ in \eqref{action} by redefining the momentum to avoid the (obvious) non-adiabatic limit caused by the vanishing momentum.  The parameter $\vec{p}_{F}$ is an arbitrary parameter, which does not exist in the action of the original literature~\cite{yamamoto, niu, duval}, and it does not necessarily implies the Fermi momentum $\vec{p}_{F}$. To ensure the true adiabatic limit, one needs to constrain the time dependence of dynamical variables also ("adiabatic" implies time-dependence), but it is not done in a simple manner. 

The applicability of this action \eqref{action} beyond the precise adiabatic movement of dynamical variables is not established, but we tentatively assume that  this action is used for the general movement of the electron to evaluate equal-time commutation relations. It is also shown later that this action is non-local in time (contains infinitely higher derivative terms) and thus no canonical quantization is applicable.  

We next observe that  canonical commutation relations are not modified by any static potential in quantum mechanics.
Only the terms with explicit time derivative such as the kinetic energy term are essential, and  the absolute size of the potential such as the harmonic oscillator potential does not matter. We can use the quadratic expansion of the Lagrangian around the origin of the space of dynamical variables to analyze the commutation relations.  
We thus consider the quadratic form of \eqref{action} around $(\vec{x},\vec{p})=(0,\vec{p}_{F})$ by redefining the momentum $(\vec{p}-\vec{p}_{F})\rightarrow \vec{p}$,
\begin{eqnarray}\label{action2}
S=\int dt[p_{k}\dot{x}_{k} +\frac{1}{2}F_{lk}x_{l}\dot{x}_{k} - \frac{1}{2}\Omega_{lk}p_{l}\dot{p}_{k} -\frac{\vec{p}^{2}}{2m}]
\end{eqnarray}
where $F_{lk}=\partial_{l}A_{k}(0)-\partial_{k}A_{l}(0)$ and $\Omega_{lk}=\partial_{l} {\cal A}_{k}(\vec{p}_{F})-\partial_{k} {\cal A}_{l}(\vec{p}_{F})$.
One can then confirm the basic correlation function  by first integrating over $\vec{p}$ in the path integral $\int{\cal D}\vec{x}{\cal D}\vec{p}\ x_{k}(t)x_{l}(0)\exp[(i/\hbar)S]$ as,
\begin{eqnarray}\label{propagator}
&&\int dt e^{i\omega t}\langle T^{\star} x_{k}(t)x_{l}(0)\rangle\nonumber\\
&&=i\hbar\{[1+im\omega\Omega]^{-1}m\omega^{2}+i\omega F\}^{-1}\nonumber\\
&&= \frac{i\hbar}{m\omega^{2}+i\omega F -m\omega^{2}\Omega F}(1+im\omega \Omega),
\end{eqnarray}
where it is understood that $kl$ matrix element of the right-hand side is taken  in the following, together with the equation of motion
$p_{k}=m[\delta_{kl}-(\Omega F)_{kl}]\dot{x}_{l}$.

The correlation \eqref{propagator} vanishes for $\omega\rightarrow \infty$ and thus we can replace $T^{\star}$ by the canonical $T$. We then multiply the both hand side by $-i\omega$ and consider the large $\omega$ limit.
We then obtain 
\begin{eqnarray}\label{propagator2}
&&\lim_{\omega\rightarrow\infty}\int dt e^{i\omega t}\{\langle T \dot{x}_{k}(t)x_{l}(0)\rangle+\delta(t)[x_{k}(0),x_{l}(0)]\}\nonumber\\
&&=i\hbar\frac{1}{1-\Omega F}\Omega
\end{eqnarray}
where we used $\frac{d}{dt}\langle T x_{k}(t)x_{l}(0)\rangle=\langle T \dot{x}_{k}(t)x_{l}(0)\rangle+\delta(t)[x_{k}(0),x_{l}(0)]$. We thus conclude 
\begin{eqnarray}\label{commutator1}
[x_{k}(0),x_{l}(0)]=i\hbar\frac{1}{1-\Omega F}\Omega
\end{eqnarray}
and 
\begin{eqnarray}\label{propagator3}
&&\int dt e^{i\omega t}\langle T \dot{x}_{k}(t)x_{l}(0)\rangle\\
&=& \frac{\hbar\omega}{m\omega^{2}+i\omega F -m\omega^{2}\Omega F}(1+im\omega \Omega)-i\hbar\frac{\Omega}{1-\Omega F}.\nonumber
\end{eqnarray}
We then multiply both hand sides of this relation by $-i\omega$ and examine the behavior for $\omega\rightarrow \infty$. Repeating this procedure we obtain
\begin{eqnarray}\label{commutator2}
&&[\dot{x}_{k}(0),x_{l}(0)]=-i\hbar[\frac{1/m}{1-\Omega F}+\frac{1/m}{1-\Omega F}F\frac{1}{1-\Omega F}\Omega],\nonumber\\
&&[\dot{x}_{k}(0),\dot{x}_{l}(0)]=-i\hbar[\frac{1/m}{1-\Omega F}F\frac{1/m}{1-\Omega F}\nonumber\\
&&\hspace{2.3cm}+\frac{1/m}{1-\Omega F}F\frac{1/m}{1-\Omega F}F\frac{1}{1-\Omega F}\Omega],
\end{eqnarray}
and $[\ddot{x}_{k}(0),x_{l}(0)]=-[\dot{x}_{k}(0),\dot{x}_{l}(0)]$.
We thus reproduce the quantized version of Poisson brackets  in \cite{yamamoto, duval} except for the last term in $[\dot{x}_{k}(0),\dot{x}_{l}(0)]$. This agreement implies that the Poisson bracket implicitly treats the variables with time derivatives preferentially, irrespective of the size of the potential term. More importantly, the agreement of the two schemes (to the order relevant to the analysis in the present paper) shows that both schemes use the behavior of dynamical variables in the extremely non-adiabatic region with $\omega\rightarrow \infty$ in an essential way, for which the derivation of the effective action \eqref{action} completely fails. Here we briefly comment on what we mean by high frequency region.  When one defines
\begin{eqnarray}
{\cal F}(\omega)=\int dt e^{i\omega t}\langle 0|T x_{k}(t)x_{l}(0)|0\rangle
\end{eqnarray}
we confirmed ${\cal F}(\infty)=0$, but $\lim_{\omega\rightarrow\infty}\omega{\cal F}(\omega)\sim \langle 0|[x_{k},x_{l} ]|0\rangle=\sum_{n}\left(\langle 0|x_{k}|n\rangle\langle n|x_{l}|0\rangle-\langle 0|x_{l}|n\rangle\langle n|x_{k}|0\rangle\right)$, namely, all the intermediate states contribute with equal weight.  If one cuts off the frequency such as ${\cal F}(\omega)\sim \exp[-\omega^{2}/\omega_{c}^{2}]$, for example, no non-trivial equal-time commutators appear.  This is an ingenious insight of BJL.

The denominator of the basic correlation function \eqref{propagator} is written as 
$(1-im\omega\Omega +(im\omega\Omega)^{2} - .....)m\omega^{2} + i\omega F$,
namely, it contains an infinite series in $\omega$ which shows that the theory is non-local in time and thus no canonical quantization is possible~\cite{non-commutative}, although the action \eqref{action} is formally written in terms of $\vec{x}$ and $\vec{p}$ and thus looks superficially canonical; it is no more canonical if one adds a point-like monopole at the origin of the momentum space.  The path integral defines a correlation function even for a theory non-local in time but one cannot convert it to canonical formulation. A consequence of this non-locality is that we have an infinite tower of commutation relations in \eqref{commutator2} containing arbitrary higher time derivative terms. 
The assumption of a point-like monopole in momentum space, which gives rise to the correlation \eqref{propagator}, leads to anomalous space-time commutation relations similar to the non-commutative space-time~\cite{non-commutative}. This clear recognition of non-locality is an advantage of the BJL method.

We recall that the monopole-type structure of Berry's phase is valid only in the precise adiabatic limit $\eta=2v_{F}|\vec{p}| /\hbar\omega\rightarrow\infty$~\cite{simon}. Only in this limit we can have (for the half-sphere)
\begin{eqnarray}\label{flux}
\oint{\cal A}_{k}(\vec{p})\dot{p}_{k}dt=\int\Omega dS =\pi
\end{eqnarray}
and this value of $\Omega$ appearing in the commutator \eqref{commutator1} determines the coefficient of "chiral anomaly" in the derivation of \cite{yamamoto}; any deviation from the monopole value would invalidate the derivation of anomaly.
Since Hamiltonian \eqref{hamiltonian} is valid only in the vicinity of the crossing point of two levels in the band structure, $|\vec{p}|$ is much smaller than $|\vec{p}_{F}|$ and thus $\omega$ contained in $\vec{p}(t)$ is required to be very small to satisfy adiabaticity.  Note that for any value 
$|\vec{p}| \leq |\vec{p}_{F}|$,
\begin{eqnarray}
2v_{F}|\vec{p}| /\hbar\omega\leq \eta =2v_{F}|\vec{p}_{F}| /\hbar\omega \rightarrow 0
\end{eqnarray}
in the limit $\omega\rightarrow \infty$, and this momentum value belongs to the non-adiabatic domain in the analysis of commutators. 
The point-type monopole structure of $\Omega$ we assumed in \eqref{action2}, which always satisfies $\int\Omega dS =\pi$,  is not valid for large $\omega$, and it is strongly suppressed in the non-adiabatic limit with small $\eta$
typically in the form ( {\em if one uses the correct value of Berry's phase}), 
\begin{eqnarray}
\oint{\cal A}_{k}(\vec{p})\dot{p}_{k}dt=\int\Omega dS\simeq \pi\eta^{2}
\end{eqnarray}
as in \eqref{non-adiabatic} which goes as $1/\omega^{2}$ for large $\omega$.  
We tentatively assumed that \eqref{action} with a monopole at the origin of the momentum space is valid for the general class of dynamical variables and derived \eqref{commutator1} and \eqref{commutator2}. But the derivation of the effective action \eqref{action} itself completely fails for the frequency regions used in the BJL analysis.

To summarize the above analysis, one recognizes that  the direct use of Berry's phase and the  indirect use of Berry's phase through an effective action and equal-time commutators are very different:
For the {\bf direct use} of the action in \eqref{action},  one obtains an equation of motion with a monopole at the origin of the momentum space. One can choose the time dependence of dynamical variables and the value of $|p|$ 
 such that the {\em approximate} adiabatic condition is satisfied.  One can thus maintain
 $\oint{\cal A}_{k}(\vec{p})\dot{p}_{k}dt=\int\Omega dS \simeq \pi$ as in \eqref{flux}.  Namely, one can {\em choose} the systems which satisfy the adiabaticity conditions approximately~\cite{niu}. To my knowledge, all the successful applications of Berry's phase in the past are the direct use of the action. But the model with a pure monopole \eqref{action} totally fails to describe Berry's phase off the adiabatic limit implied by the exact solution of Berry's model analyzed in section 2. 
 
For the {\bf indirect use} of the action \eqref{action2} to derive the anomalous commutators, 
one finds that the identical form of anomalous commutators with the monopole value $\int\Omega dS=\pi$ appears irrespective of the value of $p_{F}$ even for $p_{F}\sim 0$.  Mathematically, the extremely non-adiabatic limit $\omega\rightarrow \infty$ are important to determine the commutators, and in this limit the assumed point-like monopole in the momentum space controls the commutators independently of the adiabatic or non-adiabatic domain.  Once an explicit form of an effective action is written, either Poisson bracket or BJL prescription automatically determines the form of commutators. 

We want to  incorporate the applicability condition of Berry's phase faithfully in the evaluation of commutators. The only way to do so is to choose the action which is valid in the extreme non-adiabatic domain  to be consistent with BJL prescription.  This action may not be accurate in the adiabatic limit.
Since Berry's phase, when measured in a natural manner described in section 2,  vanishes for 
$\omega\rightarrow \infty$ as in \eqref{non-adiabatic2} irrespective of any finite $|p|$ or $|B|$,
we define the effective action without Berry's phase,
which gives accurate equations of motion in the nonadiabatic domain where BJL method is applicable,
\begin{eqnarray}\label{action3}
S=\int dt[p_{k}\dot{x}_{k} +A_{k}(\vec{x},t)\dot{x}_{k} -\frac{\vec{p}^{2}}{2m}].
\end{eqnarray}
This gives the standard free non-relativistic motion of the electron.  One can confirm that BJL method and Poisson brackets applied to \eqref{action3} reproduce the ordinary commutation relations for $\vec{x}$ and $\vec{p}$ which are obtained by setting $\Omega=0$ in \eqref{commutator1} and \eqref{commutator2}. This analysis confirms that no anomalous {\em equal-time} space-time commutation relations are induced by the mere presence of a ``Weyl fermion'' (such as a massless Weyl neutrino.)

Ideally, one may want to have an effective action which incorporates the exact information of Berry's phase for general movement of dynamical variables,
but it is not feasible. 
We thus chose two explicit effective actions,  the action \eqref{action} which gives a correct equation in the precise adiabatic limit with a monopole value of $\Omega$ but fails in the non-adiabatic domain, and the action \eqref{action3} which gives a correct equation in the precise non-adiabatic limit with vanishing $\Omega=0$ but fails off the exact non-adiabatic domain.  From the consistency with BJL prescription we choose the commutators given by the action \eqref{action3} as
physical ones.   

  As for the action \eqref{action}, we suggest that  it
  should be regarded  as an action for a charged particle with a monopole located at the origin of the momentum space.  ( If the action \eqref{action} or \eqref{action2} should be shown to those unfamiliar with  Berry's phase, they would recognize it as an action for a particle with a monopole placed at the origin of the momentum space {\em without} any constraint on the movement of variables. )
The appearance of anomalous space-time commutators in \cite{duval, yamamoto} is understood as an artifact that they used an effective action \eqref{action} or \eqref{action2} which describes  a genuine point-like monopole placed at the origin of the momentum space. It is well-known that the magnetic field modifies equal-time commutators such as in the Landau level.

In conclusion, we choose the action \eqref{action3} which is consistent with the exact solution of Berry's model and the BJL analysis as a prediction of Berry's phase to evaluate equal-time commutators, and we obtain no anomalous space-time commutators.
Physicists tend to assume the monopole form of Berry's phase everywhere, which is actually valid only in the precise adiabatic limit~\cite{simon, fuji-deguchi}.
The analysis of Berry's phase in the non-adiabatic and non-topological domain 
is very common in chemistry~\cite{longuet-higgins}. One may recall that the treatment of molecular systems using the Born-Oppenheimer approximation led to the original discovery of the geometric phase in the form of Longuet-Higgins phase change rule~\cite{longuet-higgins} in chemistry.

\section{Chiral anomaly from a monopole}
The mere presence of a Weyl fermion does not induce non-commutative space-time and thus no anomaly is induced by Berry's phase. However, a genuine point-like monopole placed at the origin of momentum space induces non-commutative space-time,
as we have analyzed above in \eqref{commutator1} and \eqref{commutator2}, in agreement with  the analysis in \cite{duval, yamamoto}.
Based on this latter assumption, the authors of \cite{yamamoto} discussed the anomalous equal-time commutation relations of the charge density operator $j^{0}=n(x)$ induced by the anomalous space-time commutators. They then derived a ``kinematic'' relation which formally agrees with the covariant chiral anomaly by an analysis of $[j^{0}, H]$.

Here, we discuss the mathematical consistency of their formulation, since this problem is analogous to the well-known interesting but controversial problem of an interplay of monopole and anomaly in the fermion-monopole system in the presence of  a genuine point-like monopole (Callan-Rubakov effect)~\cite{yamagishi}. It is interesting to see if a monopole at the origin of the momentum space gives a flux corresponding to the  correct chiral anomaly.

Since they derived the possible anomaly using equal-time commutation relations of currents \cite{yamamoto}, we first briefly summarize the known basic properties of equal time commutators associated with the conventional formulation of chiral anomaly.  
To avoid the Schwinger term we usually use the Gauss-law
operator defined by 
\begin{eqnarray}
G=j^{0}-\partial_{k}\dot{A}^{k}
\end{eqnarray}
of  chiral Abelian gauge theory in the gauge $A_{0}=0$. 
First of all, no anomalous equal-time commutation relation of the Gauss operator for Abelian chiral gauge theory is known in conventional formulation~\cite{jackiw, adler, faddeev, jo, fujikawa2}, namely, 
\begin{eqnarray}
[j^{0}(t,\vec{x})-\partial_{k}\dot{A}^{k}(t,\vec{x}), j^{0}(t,\vec{y})-\partial_{k}\dot{A}^{k}(t,\vec{y})]=0.
\end{eqnarray}
The anomalous commutation relation of the Gauss operator, which is closely related to the commutator in \cite{yamamoto},
\begin{eqnarray}\label{gauss}
\partial_{t}G(t,\vec{x})=\frac{i}{\hbar}[H, G(t,\vec{x})]=-\frac{1}{3}\frac{1}{4\pi^{2}}\vec{E}\cdot \vec{B}
\end{eqnarray}     
has been discussed in~\cite{ fujikawa-suzuki, fujikawa2}.  This relation shows that the Gauss operator, which is the generator of time independent gauge transformation,  is time dependent for chiral Abelian gauge theory and that Hamiltonian $H$ is gauge non-invariant, and thus theory is {\em inconsistent}
as is well-known. To have a consistent theory in continuum, one needs to have a vector-like theory such as the conventional QED if one does not increase the number of fermion species. 
The relation \eqref{gauss} is reduced to 
\begin{eqnarray}\label{consistent-anomaly}
\partial_{\mu}j^{\mu}=-(\frac{1}{3})\frac{1}{4\pi^{2}}\vec{E}\cdot\vec{B}
\end{eqnarray}
if one uses the equation of motion for $A^{\mu}$. The extra factor $1/3$ (and $-1/2$ due to chiral projection) 
compared to the standard form of anomaly \eqref{identity1} with $m=0$
is a characteristic of {\em consistent anomaly}. This difference arises from the fact that $j^{\mu}$ in the standard anomaly \eqref{identity1} does not couple to gauge field $A^{\mu}$; one can thus impose gauge invariance for currents appearing inside Hamiltonian (or Lagrangian) and collect the non-conservation (anomaly) only to the external non-gauge current. 
In contrast, the current appearing in Gauss-law operator in \eqref{gauss} is generated by a variational derivative of the action with respect to $A^{\mu}$ and thus a preferential treatment of one of currents is not allowed. 

More precisely, when one evaluates the anomaly by equal-time commutators, one generally encounters the commutator of the form
\begin{eqnarray}\label{current-algebra}
{[}j^{\mu}(t,\vec{x}), j^{\nu}(t,\vec{y}){]}.
\end{eqnarray}
From this commutator which treats the two currents on equal footing, one immediately recognizes that it is impossible to impose gauge invariance on one of the currents and collect the non-conservation (anomaly) to the other. This is the reason why only the consistent form of anomalies appeared in the analysis of equal-time commutation relations in the past~\cite{faddeev, jo, fujikawa2}.   This is the mathematical consistency condition nothing to do with physics.  
From a point of view of Feynman diagrams,  this factor $1/3$ in the present Abelian gauge theory is also understood as a Bose symmetrization factor of a triangle diagram, which means all the three currents are treated on equal footing.  

Coming back to the analysis in \cite{yamamoto}, the authors argued for the derivation of a ``kinematic'' relation by an analysis of $[j^{0}, H]$ with 
$j^{0}(x)=n(x)$. 
To be more explicit, they first evaluate 
\begin{eqnarray}
[n(\vec{x}), n(\vec{y})]=-i\left(\vec{\nabla}\times \vec{\sigma}+\frac{k}{4\pi^{2}}\vec{B}\right)\cdot \vec{\nabla}\delta(\vec{x}-\vec{y})
\end{eqnarray}
using the anomalous space-time commutators; the first term with $\vec{\sigma}$
arises form \eqref{commutator1} and the second term with $\vec{B}$ arises from 
\eqref{commutator2}, namely, $[\dot{x}_{k}(0),x_{l}(0)]$ combined with $p_{k}=m[\delta_{kl}-(\Omega F)_{kl}]\dot{x}_{l}$. The parameter $k=(1/2\pi)\int d\vec{S}\cdot \vec{\Omega}$ corresponds to a monopole located at the origin of the momentum space ($k=1$ for the full sphere integral in \eqref{flux}).  
They then assume the form of the Hamiltonian 
\begin{eqnarray}\label{hamiltonian-prime}
H^{\prime}= H + \int d^{3}x \phi(\vec{x})n(\vec{x})
\end{eqnarray}
with $\vec{E}=-\vec{\nabla}\phi(\vec{x})$. $H$ has a rather involved expression but not relevant for the evaluation of anomaly.
They then evaluate 
\begin{eqnarray}
\partial_{t}n(\vec{x})&=&i[n(\vec{x}),H^{\prime}]\nonumber\\
&=&-\vec{\nabla}\cdot \vec{j}+\left(\vec{\nabla}\times \vec{\sigma}+\frac{k}{4\pi^{2}}\vec{B}\right)\cdot\vec{E}\nonumber\\
&=&-\vec{\nabla}\cdot \vec{j}^{\prime}+\frac{k}{4\pi^{2}}\vec{B}\cdot\vec{E}.
\end{eqnarray}
with $\vec{j}^{\prime}=\vec{j}+\vec{E}\times \vec{\sigma}$. The relation  
\begin{eqnarray}
\partial_{t}n(\vec{x})+\vec{\nabla}\cdot \vec{j}^{\prime}=\frac{k}{4\pi^{2}}\vec{B}\cdot\vec{E}
\end{eqnarray}
is then identified with the anomalous identity with a {\em covariant anomaly}. 
The choice $k=-1$ corresponds to the left-handed convention we used in this paper so far,
\begin{eqnarray}\label{covariant-anomaly}
\partial_{\mu}j^{\mu}=-\frac{1}{4\pi^{2}}\vec{E}\cdot\vec{B}
\end{eqnarray}
for the chiral current $j^{\mu}=\bar{\psi}\gamma^{\mu}[(1-\gamma_{5})/2]\psi$, which is obtained from \eqref{identity1} combined with $\partial_{\mu}(\bar{\psi}\gamma^{\mu}\psi)=0$ if one sets $m=0$.

The essential commutator in the analysis in \cite{yamamoto} is thus given by (with $j^{0}(x)=n(x)$)
\begin{eqnarray}
{[}j^{0}(t,\vec{x}), j^{0}(t,\vec{y}){]}.
\end{eqnarray}
As we have already shown in \eqref{current-algebra}, it is mathematically impossible to treat two currents on {\em un-equal} footing in the evaluation of the equal-time commutator.
The appearance of the covariant anomaly in the evaluation of commutator is a mathematical inconsistency.
One might still argue that the derivation of anomaly in \cite{yamamoto}
is different from the conventional one, and all the currents involved can have covariant anomalies. Even in this case, all the currents including 
$j^{0}(x)=n(x)$ inside the Hamiltonian \eqref{hamiltonian-prime} are anomalous (not conserved) and thus their Hamiltonian is gauge non-invariant, and the theory is physically inconsistent just as \eqref{gauss}. 

One may conclude that a covariant form of anomaly appears in the gauge current  in the derivation of chiral anomaly from a point-like monopole located at the origin of momentum space.
\\

In passing, we comment on the evaluation of the conventional gauge theoretical 
anomalies by emphasizing the adiabatic treatment and thus compared with the idea of adiabatic phases~\cite{nair, sonoda, niemi}. My understanding of this approach, although very interesting by giving a novel physical picture, is that it illustrates that chiral anomaly in gauge field theory is evaluated using various formulations of field theory, both in Lagrangian and Hamiltonian formalism. The actual evaluation of chiral anomaly itself in \cite{sonoda}, for example, is identical to the conventional field theoretical evaluation. Also, it appears to be not easy to understand the fact that the local form of the basic identity \eqref{identity1} is valid even in a curved space-time from a purely adiabatic picture.
 
Similarity of topological properties appearing in chiral anomaly and Berry's phase have been discussed in \cite{nelson} and \cite{stone}. The Hamiltonian formalism initiated by Nelson and Alvarez-Gaume \cite{nelson} was very influential. The analysis of topology in \cite{nelson} (in particular, of $SU(2)$ anomaly) has been examined in great detail in~\cite{fujikawa4}  and the conclusion is~\cite{fujikawa4}: In the fundamental level, the difference between the two notions, anomaly and Berry's phase, is stated as follows; the topology of given gauge fields leads to level crossing in the
fermionic sector in the case of chiral anomaly and the inevitable failure of the adiabatic approximation is essential to establish the existence of anomaly, whereas the level crossing in the matter sector leads to the topology of Berry's phase only when the precise adiabatic approximation holds. These two adiabatic conditions are not compatible. The analysis of anomaly in Nelson and Alvarez-Gaume \cite{nelson} is perfectly valid without referring to Berry's phase.

\section{Chiral symmetry and species doubling}

We would like to analyze the species doublers, which are commonly associated with an assertion of a pair-wise appearance of level crossings in the Brillouin zone. 
In the recent discussions of ``Weyl fermions''~\cite{fukushima, zyuzin, hosur} and also in the earlier work~\cite{nielsen2}, 
the  chiral anomaly corresponding to \eqref{covariant-anomaly} is assumed  for a Weyl fermion at {\em each} level-crossing point such as \eqref{hamiltonian} in the band structure by identifying it as a species doubler. 

The fermion theory defined on a lattice with a finite spacing $a$ removes the entire short distances~\cite{wilson} and thus high frequencies~\cite{fujikawa-suzuki} that are responsible for the anomaly. This fact is reflected in  the appearance of species doublers in momentum space which ensure the absence of anomaly even in the limit $a\rightarrow 0$ for a manifestly  {\em chiral invariant} formulation~\cite{karsten, nielsen}.  This is symbolically stated as  the absence of a ``neutrino'' on the lattice~\cite{nielsen}, and the related {\em inevitable} appearance of species doublers seems to be assumed in condensed matter physics~\cite{hosur} presumably because the simplest nearest neighbor discretization of a massless fermion in the tight-binding approximation (latticed model) suggests such behavior. 

The notion of species doublers means that a single ``local'' fermion and thus a single ``local'' current defined on the lattice actually describes the multiple species of fermions, i.e., doublers,  in momentum space in the limit $a\rightarrow 0$. 
For example, if one attempts to define a left-handed massless fermion
\begin{eqnarray}
\psi_{L}(x)=[(1-\gamma_{5})/2]\psi(x)
\end{eqnarray}
on the lattice, the species doubling implies that one inevitably obtains both a left-handed fermion and a right-handed fermion in momentum space in the limit $a\rightarrow 0$.  Similarly, $\psi_{R}(x)=[(1+\gamma_{5})/2]\psi(x)$ implemented on the lattice inevitably induces both a left-handed fermion and a right-handed fermion in momentum space in the limit $a\rightarrow 0$.  
The chiral fermion such as $\psi_{L}(x)$, which is inflicted with chiral gauge anomaly  \eqref{gauss} in continuum theory and thus inconsistent by itself, contains no chiral gauge anomaly when placed on the lattice which cut-off the short distances.  Both $\psi_{L}$ and $\psi_{R}$ can in principle appear simultaneously; for $a\rightarrow 0$, we then have multiple species of charged fermions (such as the massless ``electron'' and  ``muon'', although we originally intended to define only the ``electron'', and thus the construction is inconsistent).  In any case, the notion of species doubling implies that we obtain at least twice as many fermions in momentum space  in the limit $a\rightarrow 0$ than the original intention.

 On the other hand, the common knowledge  in lattice gauge theory nowadays is that a  recent progress of Ginsparg-Wilson fermion allows a definition of  a single  Weyl fermion without doublers on the lattice~\cite{neuberger}. See \cite{fujikawa-suzuki} for further references. The chiral current in such a theory is {\em exponentially local} ~\cite{locality} (stated intuitively, it contains nearest neighbor couplings, next nearest neighbor couplings, and so on ... , namely, the construction is not manifestly local but actually becomes local in the limit $a\rightarrow 0$)  without any species doublers. In this formulation, exact chiral symmetry is realized by an operator that is deformed from $\gamma_{5}$ for finite $a$; one obtains a non-trivial Jacobian as a symmetry breaking factor under chiral transformation in the path integral formulation of chiral identity~\cite{luescher}, which gives a correct anomaly in the continuum limit~\cite{fujikawa-suzuki}. 
 For  small momenta close to the origin of the momentum space, the ordinary Weyl spectrum is realized, namely, one can identify a {\em low energy chiral excitation mode} on the (hypercubic) lattice without doublers.  
Thus we have {\em no more} the "no-go theorem" against a single chiral fermion on the lattice.

In condensed matter and related fields, the species doublers are treated as physical objects, in contrast to lattice gauge theory where the species doublers are unphysical nuisances. It is then interesting to ask if the absence of anomaly in latticized theory with species doublers is caused by the cancellation of well-defined anomalies produced by each doubler or none of doublers has anomalies for {\em finite} $a$. We would like to show that none of doublers has well-defined anomalies, as is expected from the fact that  each doubler is defined only in  part of the Brillouin zone of momentum space and thus not a well-defined local field.   
Instead, a pair production associated with spectral flow in the Dirac sea with finite depth takes place. We emphasize that our analysis below is based on  lattice gauge theory, although we hope that it may have some implications on condensed matter and related fields.
 
\subsection{Species doubling and spectral flow}

To understand the following discussions intuitively, it is instructive to consider a $d=1+1$ dimensional lattice fermion.  We thus consider a model Hamiltonian
\begin{eqnarray}\label{two-dimension}
H=\sigma_{3}\frac{\sin ap}{a}+\sigma_{1}\frac{r}{a}(1-\cos ap)
\end{eqnarray}
with a constant $r$ which is usually called the Wilson parameter. Note that $H(p)$ has a period $2\pi/a$ in $p$. If one recalls that chiral matrix is given by $\gamma_{5}=\sigma_{3}$ in this notation,
the first term is chiral invariant but the second term does not commute with $\gamma_{5}$ and thus breaks chiral symmetry. The chiral symmetric Hamiltonian is thus given by setting $r=0$,
\begin{eqnarray}
H_{0}=\sigma_{3}\frac{\sin ap}{a}.
\end{eqnarray} 
We have the energy spectrum of $H_{0}$ as
\begin{eqnarray}
\epsilon^{(0)}_{\pm}(p)=\pm \frac{\sin ap}{a}
\end{eqnarray} 
where $\epsilon^{(0)}_{\pm}(p)$ correspond to chirality $\gamma_{5}=\pm 1$, respectively.  Note that $\epsilon^{(0)}_{\pm}(p+2\pi/a)=\epsilon^{(0)}_{\pm}(p)$. This exhibits the spectrum similar to the Weyl fermion for small $|\epsilon^{(0)}_{\pm}(p)|$, and $\epsilon^{(0)}_{\pm}(p)=0$ at $p=0$ and $p=\pi/a$ in the Brillouin zone. Moreover, $\epsilon^{(0)}_{+}(p)$ near $p=\pi/a$ has the same  structure as $\epsilon^{(0)}_{-}(p)$ near $p=0$, and similarly starting with $\epsilon^{(0)}_{-}(p)$. Thus we have effectively two Weyl fermions with $\gamma_{5}=1$ and two Weyl fermions with $\gamma_{5}=-1$, namely, we have {\em species doubling} of the Weyl fermion for chiral invariant theory in the limit $a\rightarrow 0$. 

We may write the solution corresponding to $\epsilon^{(0)}_{-}(p)$ in the form
\begin{eqnarray}\label{doublers}
&&\psi_{L}(x)\nonumber\\
&&=\int_{-\pi/2a}^{\pi/2a}\frac{dp}{(2\pi)}e^{-ipx}\psi_{L}(p)+\int_{\pi/2a}^{3\pi/2a}\frac{dp}{(2\pi)}e^{-ipx}\psi_{L}(p)\nonumber\\
&&=\int_{-\pi/2a}^{\pi/2a}\frac{dp}{(2\pi)}e^{-i\epsilon^{(0)}_{-}(p)t+ipx^{1}}\psi_{L}(p)\nonumber\\
&&+e^{i\pi x^{1}/a}\int_{-\pi/2a}^{\pi/2a}\frac{dp}{(2\pi)}e^{-i\epsilon^{(0)}_{+}(p)t+ipx^{1}}\psi_{L}(p+\pi/a)\nonumber\\
&&\equiv e_{L}(x)+e^{i\pi x^{1}/a}\mu_{R}(x)
\end{eqnarray}
by choosing the Brillouin zone $-\pi/2a\leq p<3\pi/a$. We defined formally two fields $e_{L}(x)$ and $\mu_{R}(x)$ although they are actually part of a single field $\psi_{L}(x)$; when one discusses short distance properties  such as the chiral anomaly of $\psi_{L}(x)$, which imply the maximum extension in momentum space by uncertainty principle,  one cannot separate $e_{L}(x)$ and $\mu_{R}(x)$. We emphasize that $e_{L}(x)$ and $\mu_{R}(x)$ separately cannot define  well-defined fields for $a\neq 0$ since half of the Brillouin zone is missing in them. Nevertheless, this notation is useful to understand the following discussions. 

For example, in the picture of the spectral flow of $\epsilon^{(0)}_{-}(p)=-\sin ap/a$ with all the negative energy states $0\leq p \leq \pi/a$ being filled initially, a particle creation at the momentum close to $p=0$ implies a hole creation close to $p=\pi/a$, namely, we have an inevitable pair production
\begin{eqnarray}\label{d=2-pairproduction}
\psi_{L}+ \bar{\psi}_{L}\ \ \ {\rm or}\ \ \ e_{L}+ \bar{\mu}_{R}.
\end{eqnarray}

For $r\neq 0$ in \eqref{two-dimension}, we have no more chiral symmetry generated by $\gamma_{5}$. We have only a single massless solution, since $H(p)=0$ means two linearly independent terms of $H(p)$ in \eqref{two-dimension} vanish simultaneously. To be explicit, we have two eigenvalues
\begin{eqnarray}\label{2d-Dirac}
\epsilon_{\pm}(p)=\pm \sqrt{\left(\frac{\sin ap}{a}\right)^{2}+ \left(\frac{r}{a}(1-\cos ap)\right)^{2}}
\end{eqnarray}
that show $\epsilon_{\pm}(p+2\pi/a)=\epsilon_{\pm}(p)$. We have $\epsilon_{\pm}(p)=\pm |p|$ near $p=0$ and $\epsilon_{\pm}(p)=\pm \frac{2r}{a}$ near $p=\pi/a$, namely, no more species doubling for $r/a\gg 1$.   We thus have a single massless Dirac fermion, but the drawback of this construction is that the chiral symmetry of Hamiltonian is completely lost. The recent progress in the Ginsparg-Wilson fermion is that one can now construct a fermion model that defines a single massless Dirac  fermion  with an exact chiral symmetry without doublers in the limit $a\rightarrow 0$.       
                              
  In the sequel, we discuss the chiral anomaly using two pictures of ``Weyl fermions'', namely, species doublers and the Ginsparg-Wilson fermion.  We first discuss the chiral anomaly using species doublers  and then  the Ginsparg-Wilson fermion to support our view.  Our analysis is performed in the framework of lattice gauge theory without referring to the details of specific fields such as condensed matter and nuclear physics.
 
\subsection{Chiral anomaly and spectral flow}

We generalize the Hamiltonian \eqref{hamiltonian} to a full-fledged Weyl-type wave equation
\begin{eqnarray}\label{weyl-equation}
[i\frac{\partial}{\partial t}-v_{F}\vec{\sigma}\cdot\frac{1}{i}\frac{\partial}{\partial\vec{x}}]u(t,\vec{x})=0
\end{eqnarray} 
where $v_{F}>0$ or $v_{F}<0$ corresponds to right-handed or left-handed fermions, respectively. We set $\hbar=1$ for simplicity.
Since this equation is defined as an effective equation in the Brillouin zone with $a\neq 0$, we attempt to {\em simulate} this equation by an effective lattice gauge theory treating the pseudo-spin $\vec{\sigma}$ as a real spin.  
The purpose of this simulation is to obtain possible new insights into condensed matter and related fields with regards to quantum anomaly.

We thus start with a massless Dirac fermion in continuum
\begin{eqnarray}\label{continuum-Dirac}
i\gamma^{\mu}(\partial_{\mu}-iA_{\mu})\psi=0
\end{eqnarray}
with the electromagnetic gauge field $A_{\mu}$. We then put this equation on
the hypercubic lattice, for example, with manifest chiral symmetry generated 
by $\gamma_{5}$. There are two ways to latticize the continuum gauge theory.
The first one is the Hamiltonian formalism that exhibits the energy-momentum dispersion in the band diagram nicely. In this scheme, it is known that the minimum number of species doublers appear~\cite{nielsen2}, namely, one species doubler for each chiral component that are sufficient to ensure the absence of chiral anomaly even in the limit $a\rightarrow 0$. The other formulation is the Euclidean Lagrangian formalism, which generally contains more species doublers than the Hamiltonian formulation, but the analysis of chiral anomaly is more transparent. Also, it is this Euclidean formulation that led to a formulation of the Ginsparg-Wilson fermion containing no species doublers by deforming the generator of chiral symmetry from $\gamma_{5}$.  

We here use the $d=4$ Euclidean formulation, although the physical contents in the band diagram of condensed matter theory are not directly seen in this formulation (to see the band structure, we need to consider the domain of small energy-momentum combined with Wick rotation, and then \eqref{continuum-Dirac} is realized).
In this notation, we have (see \eqref{doublers}),
\begin{eqnarray}\label{fermion1}
\psi(x)=e(x)+ \gamma_{3}\gamma_{5}e^{i\pi x_{3}/a}\mu(x)
\end{eqnarray}   
where $e(x)$ and $\mu(x)$ are "Dirac fermions" defined in the  momentum domain  $-\pi/2a\leq k_{\mu}<\pi/2a$ when the Brillouin zone is chosen as $-\pi/2a\leq k_{\mu}<3\pi/2a$. We emphasize that we have actually only a single physical field $\psi(x)$ defined in the entire Brillouin zone; see \eqref{doublers} for a related discussion. Two massless Dirac fermions $e(x)$ and $\mu(x)$ are displaced on the lattice in the $z$-direction by $\pi/a$. We have actually more species doublers in 
$d=4$ Euclidean formulation (due to the time discretization and also the hypercubic symmetry), but the essential aspect we want to discuss is correctly captured by this simplified model \eqref{fermion1}. 

By multiplying the chiral projectors, we find from \eqref{fermion1}
\begin{eqnarray}\label{d=4-chiral-fermion}
\psi_{L,R}(x)&=&(\frac{1\mp \gamma_{5}}{2})\psi(x)\nonumber\\
&=&e_{L,R}(x)+ \gamma_{3}\gamma_{5}e^{i\pi x_{3}/a}\mu_{R,L}(x)
\end{eqnarray}
This shows that the left-handed fermion $\psi_{L}(x)$, for example, when placed on the lattice becomes a superposition of 
two  fermions $e_{L}(x)$ and $\mu_{R}(x)$. Two  fermions are actually part of the single field $\psi_{L}(x)$ although they are displaced in momentum space.  
In the limit $1/a \rightarrow  {\rm large}$, the coherence between $e_{L}(x)$ and $\mu_{R}(x)$  is lost (and quantum tunneling between these two states in momentum space is suppressed), and thus two fermions behave approximately as two ``Weyl fermions''.  We have approximately
\begin{eqnarray}\label{left-right-fermions}
S&=&\bar{\psi}_{L}D \psi_{L}\nonumber\\
&\simeq&\bar{e}_{L}D e_{L}+\bar{\mu}_{R}D \mu_{R}
\end{eqnarray}
and similarly for $\psi_{R}(x)$. Note that the local chiral current in coordinate space carries no anomaly for $a\neq 0$~\cite{karsten, nielsen},  
$\partial_{\mu}(\bar{\psi}_{L}\gamma^{\mu}\psi_{L})(x)=0$. 
The  semi-classical spectral flow on the basis of $\psi_{L}$ in \eqref{d=4-chiral-fermion} shows that only the pair production from the vacuum such as 
\begin{eqnarray}\label{pair-production}
\psi_{L}+ \bar{\psi}_{L}\ \ \ {\rm or}\ \ \ e_{L}+ \bar{\mu}_{R}
\end{eqnarray}
is allowed. See the detailed discussion in the context of \eqref{d=2-pairproduction}. Thus the selection rule $\Delta N_{\psi_{L}}=0$ or
\begin{eqnarray}\label{selection-rule}
\Delta (N_{e_{L}}+ N_{\mu_{R}})=0
\end{eqnarray}
is satisfied, although two (would-be) Weyl fermions $e_{L}(x)$ and $\mu_{R}(x)$ are far apart by
\begin{eqnarray}\label{pair-production}
\Delta p \simeq \pi/a
\end{eqnarray}
in momentum space for $1/a \rightarrow {\rm large}$. 
 The analysis of possible anomaly produced by each of Weyl fermions $e_{L}(x)$ and $\mu_{R}(x)$ separately is not needed in this pair creation other than the anomaly-free conservation law $\partial_{\mu}(\bar{\psi}_{L}\gamma^{\mu}\psi_{L})(x)=0$.

In some applications, however, one may want to know an explicit form of the (possible) anomaly for each ``Weyl fermion'' separately~\cite{zyuzin, hosur, nielsen2}, but the chiral anomaly for each doubler  with $\gamma_{5}$ as the generator of chiral symmetry generally
vanishes for $a\neq 0$. This is understood by recalling the fact that the anomaly as a Jacobian (a local version of index) is expressed using a general form of the chiral invariant lattice Dirac operator $D$~\cite{fujikawa-suzuki},
\begin{eqnarray}\label{chiral-anomaly}
\lim_{M\rightarrow \infty}\int_{B} \frac{d^{4}p}{(2\pi)^{4}}{\rm tr}\{\gamma_{5}e^{-ipx}\exp[-\frac{D^{\dagger}D}{M^{2}}]e^{ipx}\}=0
\end{eqnarray}  
for a fixed finite lattice spacing $a$. Here $B$ is the Brillouin zone (or any sub-domain), of which volume is finite for finite $a$.   
 This vanishing of anomaly is a local version of the vanishing index
\begin{eqnarray} 
{\rm Tr}\gamma_{5}=0
\end{eqnarray}
which holds very generally on the lattice. See eq.\eqref{sumrule} for the Ginsparg-Wilson fermion in the next section.
Only after taking the precise limit $a\rightarrow 0$ in \eqref{chiral-anomaly} first, for which each species doubler is treated as a full fledged field, one can define the ordinary chiral anomaly for each species doubler separately by suitably choosing the  sub-domains of  $B$.  

The absence of anomaly for each doubler separately with a lattice cut-off and thus with a finite number of degrees of freedom is also understood by an argument of spectral flow by considering the simple example \eqref{doublers}. Let's concentrate on the field $e_{L}(x)$ that has the spectrum $\epsilon_{-}(p)=-\sin ap/a$ defined only in half of the Brillouin zone 
\begin{eqnarray}
-\pi/2a < p < \pi/2a, 
\end{eqnarray}
although precisely speaking no physical field on the lattice with $a\neq 0$ is defined in the domain limited to $|p|\leq \pi/2a$. If one considers a particle production near the Fermi level $\epsilon_{-}(0)=0$, a hole is generated at the lowest end of the spectrum in the deep inside the Dirac sea with energy $\epsilon_{-}(\pi/2a)=-1/a$ at $p=\pi/2a$ by the spectral flow that preserves the total fermion number. Thus a particle-hole pair creation 
\begin{eqnarray}
e_{L} + \bar{e}_{L}
\end{eqnarray}
with $\Delta N_{e_{L}}=0$ which implies $\partial_{\mu}(\bar{e}_{L}\gamma^{\mu}e_{L})(x)=0$ rather than a net particle production (i.e., anomaly) takes place.  This mechanism is the same as the pair creation in \eqref{d=2-pairproduction} although the hole is separated from the particle not only in momentum but also in energy in the present case.
  
   To have a {\em net particle production} associated with the chiral anomaly, the infinitely deep Dirac sea with an infinite number of degrees of freedom and mathematics something like $\infty+1=\infty$ is essential in the conventional picture of spectral flow. This consideration shows an important difference between anomaly and spectral flow in lattice theory which implies the Dirac sea with finite depth. It shows that  the use of an explicit form of anomaly such as  
\begin{eqnarray}\label{leftright-anomaly}
&&\partial_{\mu}(\bar{e}_{L}\gamma^{\mu}e_{L})(x)=-(1/4\pi^{2})\vec{E}\cdot\vec{B},\nonumber\\
&&\partial_{\mu}(\bar{\mu}_{R}\gamma^{\mu}\mu_{R})(x)=+(1/4\pi^{2})\vec{E}\cdot\vec{B}
\end{eqnarray}
for each species doubler separately in the picture of species doubling for $a\neq 0$ has no justifiable basis. With the help of the Ginsparg-Wilson fermion discussed in the next section that contains only a single species, one can intuitively understand that the pair production corresponds to 
${\rm Tr}\gamma_{5}=0$ in \eqref{sumrule}, while the  anomaly ${\rm Tr}\Gamma_{5}\neq 0$ in \eqref{index} in the next section projects out the hole in the deep Dirac sea. To have anomaly on the lattice, it is essential to eliminate the holes at the bottom of the energy spectrum and thus effectively realizing the infinitely deep Dirac sea. 

The difference of the presence or absence of anomaly is that we would have well-defined \eqref{leftright-anomaly} in the presence of anomaly, while in the absence of separate anomaly we have 
\begin{eqnarray}\label{leftright-anomaly2}
&&\partial_{\mu}(\bar{e}_{L}\gamma^{\mu}e_{L})(x)=0,\nonumber\\
&&\partial_{\mu}(\bar{\mu}_{R}\gamma^{\mu}\mu_{R})(x)=0.
\end{eqnarray}
In the picture of spectral flow, the first relation in \eqref{leftright-anomaly2} shows the cancellation of a particle near the fermi surface by a hole deep inside the Dirac sea, and the second relation shows the cancellation of a hole near the surface by a ``particle'' deep inside the Dirac sea.
But in both cases \eqref{leftright-anomaly} and \eqref{leftright-anomaly2}, we always satisfy 
\begin{eqnarray}\label{no-anomaly}
\partial_{\mu}(\bar{e}_{L}\gamma^{\mu}e_{L})(x)+\partial_{\mu}(\bar{\mu}_{R}\gamma^{\mu}\mu_{R})(x)=0
\end{eqnarray}
and both satisfy the fermion number conservation $\Delta N_{e_{L}}=0$. 

We finally comment on the common assertion of the inevitable pair-wise appearance of "Weyl fermions" (level crossings) in the band diagram in condensed matter physics. The chiral symmetry of the fundamental electron in the non-relativistic Schr\"{o}dinger equation determines the band structure and possible species doublers. To my knowledge, no convincing argument has been given as to the equivalence of the 
{\em pseudo-chiral symmetry} of the effective fermion in \eqref{hamiltonian} and the fundamental chiral symmetry of the electron in the Schr\"{o}dinger equation. We have shown that the spectral flow, on which the argument for the pair-wise appearance is often based, is a manifestation of the fermion number conservation which is not the same as the chiral anomaly in those theories where the ultraviolet part of the spectrum is truncated. (In the picture of a creation of fermion-antifermion pair, the hole in the deep bottom of the Dirac sea could be filled by external doping, for example, and thus the  appearance of a single isolated level crossing point in the Brillouin zone could be consistent in the picture of spectral flow which does not induce  anomaly.) It is my opinion that a clear explanation of this fundamental issue of two different chiral symmetries associated with pseudospin and real spin in the context of band theory is urgently needed.
 At this moment, it may be natural to understand that a specific configuration in the band diagram is  nicely simulated by a hypothetical lattice gauge theory with species doublers~\cite{nielsen2}.

\section{ Ginsparg-Wilson fermion}
We recapitulate some representative properties of the Ginsparg-Wilson fermion. Using the mathematically precise formulation of the Ginsparg-Wilson fermion,
we illustrate that the states at the ultraviolet region such as the states at the bottom of the Dirac sea are crucial to analyze the phenomena of anomaly and spectral flow on the lattice which has the energy bound at $\sim |1/a|$, and thus  supporting our view presented in section V. We can also understand  that the apparently low energy property such as the Atiyah-Singer index theorem is controlled by the ultraviolet cut-off in the lattice formulation. 

 The  Ginsparg-Wilson fermion is defined by 
\begin{eqnarray}\label{GWfermion}
S=\sum_{x,y}\bar{\psi}(x)D(x,y)\psi(y)
\end{eqnarray}   
on the $d=4$ hypercubic lattice, for example.
This fermion operator satisfies the Ginsparg-Wilson relation~\cite{GW}
\begin{eqnarray}\label{GWrelation}
\gamma_{5}D + D\gamma_{5} = aD\gamma_{5}D.
\end{eqnarray}  
 We then define two projection operators,
 \begin{eqnarray}\label{projectors}
P_{\pm}=\frac{1}{2}(1\pm \gamma_{5}),\ \ \ \hat{P}_{\pm}=\frac{1}{2}(1\pm \hat{\gamma}_{5})
\end{eqnarray}  
with $\hat{\gamma}_{5}=\gamma_{5}(1-aD)$ that satisfies $\hat{\gamma}_{5}^{2}=1$ using the Ginsparg-Wilson relation \eqref{GWrelation}. We define the chiral components by 
\begin{eqnarray}\label{chiral-components}
\psi_{R,L}=\hat{P}_{\pm}\psi, \ \ \bar{\psi}_{L,R}=\bar{\psi}P_{\pm}
\end{eqnarray}
which satisfy
\begin{eqnarray}\label{GWfermion2}
S=\bar{\psi}_{L}D\psi_{L}+\bar{\psi}_{R}D\psi_{R}
\end{eqnarray} 
using $D=P_{+}D\hat{P}_{-}+P_{-}D\hat{P}_{+}$. We define chiral transformation
\begin{eqnarray}\label{chiral-transformation}
\psi\rightarrow e^{i\alpha \hat{\gamma}_{5}}\psi, \ \ \bar{\psi}\rightarrow \bar{\psi}e^{i\alpha \gamma_{5}}
\end{eqnarray}  
under which the action is invariant since
\begin{eqnarray}
\gamma_{5}D+D\hat{\gamma}_{5}=0.
\end{eqnarray}
Under the transformation \eqref{chiral-transformation}, we have a Jacobian of the form
\begin{eqnarray}\label{jacobian}
\exp[-2i{\rm Tr}\alpha\frac{1}{2}(\gamma_{5}+\hat{\gamma}_{5})]=\exp[-2i{\rm Tr}\alpha\Gamma_{5}].
\end{eqnarray} 
where 
\begin{eqnarray}
\Gamma_{5}\equiv \gamma_{5}(1-\frac{1}{2}aD).
\end{eqnarray}
It is known that one can construct the operator $D$ which satisfies the Ginsparg-Wilson relation and hermiticity condition $(\gamma_{5}D)^{\dagger}=(\gamma_{5}D)$, and  contains a single species in the Brillouin zone, which is mainly concentrated  in the sub-domain $-\pi/2a\leq p_{\mu}<\pi/2a$ without any doublers~\cite{neuberger}. 

One can show~\cite{fujikawa3} that all the normalizable eigenstates
of $H\equiv \gamma_{5}D$, 
\begin{eqnarray}
H\phi_{n}=\lambda_{n}\phi_{n}
\end{eqnarray}
on the lattice with a finite spacing are categorized into the following 3 classes
using the basic relation derived from \eqref{GWrelation},
\begin{eqnarray}\label{chiral-algebra}
\Gamma_{5}H + H\Gamma_{5}=0
\end{eqnarray}
with $\Gamma_{5}=\gamma_{5}-\frac{1}{2}aH$:\\
(i) Zero modes ($n_{\pm}$ states),
\begin{eqnarray}
H\phi_{n} = 0,\  \gamma_{5}\phi_{n} = \pm \phi_{n}, 
\end{eqnarray}
(ii) Highest states ($N_{\pm}$ states) with $\Gamma_{5}\phi_{n}=0$,
\begin{eqnarray}\label{higheststates}
H\phi_{n} = \pm\frac{2}{a}\phi_{n},\ \gamma_{5}\phi_{n} = \pm \phi_{n},
\end{eqnarray}
respectively,  and \\
(iii) Remaining paired states with $0 < |\lambda_{n}|< 2/a$,
\begin{eqnarray}
H\phi_{n}=\lambda_{n}\phi_{n},\   H(\Gamma_{5}\phi_{n})=-\lambda_{n}(\Gamma_{5}\phi_{n}),
\end{eqnarray}
and the sum rule 
\begin{eqnarray}
n_{+}+N_{+}=n_{-}+N_{-}
\end{eqnarray}
holds. This sum rule is a result of
\begin{eqnarray}\label{sumrule}
{\rm Tr}\gamma_{5}=\sum_{n} (\phi_{n}, \gamma_{5}\phi_{n})=n_{+}+N_{+}-(n_{-}+N_{-})=0
\end{eqnarray}
that holds even for non-Abelian Yang-Mills fields.

The chiral symmetry breaking (non-trivial Jacobian) in \eqref{jacobian} is described by the Atiyah-Singer index theorem~\cite{fujikawa3}
\begin{eqnarray}\label{index}
{\rm Tr}\Gamma_{5}=\sum_{n} (\phi_{n}, \Gamma_{5}\phi_{n})=n_{+}-n_{-}.
\end{eqnarray}

The difference of $\Gamma_{5}$ and $\gamma_{5}$ is very important; $\Gamma_{5}$ projects out the highest states $N_{\pm}$ in \eqref{higheststates}.  The Hilbert space of continuum theory in the present scheme is defined by projecting out those $N_{\pm}$ states in the precise limit $a\rightarrow 0$.
The difference of $\Gamma_{5}$ and $\gamma_{5}$ has been already used to discuss the difference between the spectral flow and chiral anomaly in section V, where the role of the states at the bottom of the Dirac sea is identified with the role of the states $N_{\pm}$; spectral flow without anomaly corresponds to \eqref{sumrule} and the anomaly \eqref{index} is realized only when one projects out $N_{\pm}$. 
It is also interesting that the apparently low-energy statement of the index theorem is controlled by the ultraviolet cut-off if one uses 
\eqref{sumrule}, $n_{+}-n_{-}=N_{-}-N_{+}$.

The index $n_{+}-n_{-}$ is given by the Chern-Pontryagin number in the continuum limit for non-Abelian gauge theory, which is an analogue of the integral of $\frac{1}{4\pi^{2}}\vec{E}\cdot\vec{B}$. The index vanishes for Abelian theory in the absence of a magnetic monopole, but a local version of ${\rm Tr}\Gamma_{5}$ gives the correct anomaly in the continuum limit~\cite{fujikawa-suzuki},
\begin{eqnarray}\label{local-index}
&&\lim_{M\rightarrow \infty, a\rightarrow 0}\int_{B} \frac{d^{4}p}{(2\pi)^{4}}{\rm tr}\{e^{-ipx}\Gamma_{5}\exp[-\frac{(\gamma_{5}D)^{2}}{M^{2}}]e^{ipx}\}\nonumber\\
&&=\frac{1}{4\pi^{2}}\vec{E}\cdot\vec{B}.
\end{eqnarray} 
The Ginsparg-Wilson fermion is tightly constrained by the consideration of (local) index even for $a\neq 0$. 
Thus a single Weyl fermion obtained by projecting  \eqref{GWfermion} to a chiral component without doublers 
\begin{eqnarray}\label{Weyl1}
S=\bar{\psi}_{L}D\psi_{L}
\end{eqnarray} 
is not defined quantum mechanically in path integral even for $a\neq 0$ for the gauge group that contains gauge anomaly such as the $U(1)$ electromagnetism~\cite{luescher2, suzuki}. This corresponds to the inconsistency of continuum chiral gauge theory in \eqref{gauss}.  In many respects, the Ginsparg-Wilson fermion is close to the continuum fermion in Minkowski space.

\section{Discussion and conclusion}
 
We finally mention an alternative use of the anomaly  in analogy to  PCAC  (partially conserved axial-vector current hypothesis) which played an important role in the history of chiral anomaly~\cite{anomaly1,anomaly2}. This approach starts with
\begin{eqnarray}
\langle n^{\prime}|2im\bar{\psi}\gamma_{5}\psi|n\rangle\simeq-\langle n^{\prime}|\frac{e^{2}}{2\pi^{2}}\vec{E}\cdot\vec{B}|n\rangle
\end{eqnarray}
\\
implied by the basic identity \eqref{identity1} as a {\em low-energy} matrix element between two states $|n\rangle$ and $|n^{\prime}\rangle$ by preserving the fermion number $\partial_{\mu}(\bar{\psi}\gamma^{\mu}\psi)=0$.  
The basic task in this approach is to find an operator that may dominate $\langle n^{\prime}|2im\bar{\psi}\gamma_{5}\psi|n\rangle$ at low energies such as   the helicity operator $\langle n^{\prime}|i\psi^{\dagger}(\vec{x})\frac{1}{2}\vec{\sigma}\overleftrightarrow{\nabla}\psi(\vec{x})|n\rangle$ with the Pauli two-component spinor $\psi(\vec{x})$, for example. The dominating operator  is presumably related to the Coulomb interaction with surrounding particles, which is analogous to the strong interaction (QCD) in PCAC. This scheme does not induce ``fermion number non-conservation''.  It would be very interesting if one can formulate this approach in condensed matter or nuclear physics.
Historically, PCAC motivated the idea of ``anomaly-matching'' at the quark level and nucleon level calculations. 

In conclusion, we have shown that the mere presence of a Weyl fermion does not induce anomalous equal-time commutation relations of space-time variables if the nonadiabatic behavior of Berry's phase is carefully taken into account. The effective theory with a monopole placed at the origin of the momentum space, which fails to describe Berry's phase in the non-adiabatic domain,  gives rise to anomalous space-time commutators and thus chiral anomaly to the gauge current~\cite{yamamoto}.  We regard this appearance of the anomaly as an artifact of the postulated monopole and not a consequence of Berry's phase~\cite{venugopalan}.
We have also shown that the chiral anomaly vanishes for each species doubler separately in the scheme which treats species doublers in lattice theory as physical objects for a finite lattice spacing $a\neq 0$. Instead a general form of spectral flow in the Dirac sea with finite depth takes place. The Ginsparg-Wilson fermion which is free of species doubling supports this view. 

We expect that  the ideas developed in particle theory such as chiral anomaly and Ginsparg-Wilson fermions will find more interesting applications  in the fields such as condensed matter and nuclear physics. 
\\

I thank K. Fukushima for calling the work \cite{yamamoto} to my attention and for a useful comment.  I also thank H. Suzuki for a discussion on chiral anomaly.

\end{document}